\PassOptionsToPackage{table}{xcolor}
\documentclass[sigconf, noacm]{acmart}

\usepackage[ruled]{algorithm2e}   
\usepackage{natbib}

\usepackage{booktabs} 

\newcommand{\tabincell}[2]{\begin{tabular}{@{}#1@{}}#2\end{tabular}}
\setcopyright{rightsretained}

\acmDOI{10.475/123_4}

\acmISBN{123-4567-24-567/08/06}

\begin{document}
\title{
 An end-to-end Generative Retrieval Method for Sponsored Search Engine
}
 \subtitle{Decoding Efficiently into a Closed Target Domain}
%

\author{Yijiang Lian}
\affiliation{%
  \institution{Baidu}
}
\email{lianyijiang@baidu.com}

\author{Zhijie Chen}
\affiliation{%
  \institution{Baidu}
}
\email{chenzhijie01@baidu.com}

\author{Jinlong Hu}
\affiliation{%
  \institution{Baidu}
}
\email{hujinlong01@baidu.com}

\author{Kefeng Zhang}
\affiliation{%
  \institution{Baidu}
}
\email{zhangkefeng@baidu.com}

\author{Chunwei Yan}
\affiliation{%
  \institution{Baidu}
}
\email{yanchunwei@baidu.com}

\author{Muchenxuan Tong}
\authornote{Muchenxuan Tong works now at Didi Chuxing. }
\affiliation{%
  \institution{Baidu}
}
\email{demon386@gmail.com}

\author{Wenying Han}
\affiliation{%
  \institution{Baidu}
}
\email{hanwenying@baidu.com}

\author{Hanju Guan}
\affiliation{%
  \institution{Baidu}
}
\email{guanhanju@baidu.com}

\author{Ying Li}
\authornote{Ying Li works now at Ant Finance.}
\affiliation{%
  \institution{Baidu}
}
\email{leeyingxj@gmail.com}

\author{Ying Cao}
\authornote{Ying Cao works now at Microsoft.}
\affiliation{%
  \institution{Baidu}
}
\email{yincao@microsoft.com}

\author{Yang Yu}
\authornote{Yang Yu works now at aelf.io.}
\affiliation{%
  \institution{Baidu}
}
\email{I@reyoung.me}

\author{Zhigang Li}
\affiliation{%
  \institution{Baidu}
}
\email{lizhigang01@baidu.com}

\author{Xiaochun Liu}
\authornote{Corresponding author}
\affiliation{%
  \institution{Baidu}
}
\email{liuxiaochun@baidu.com}

\author{Yue Wang}
\affiliation{%
  \institution{Baidu}
}
\email{wangyue@baidu.com}

\renewcommand{\shortauthors}{Yijiang Lian et al.}

\begin{abstract}
In this paper, we present a generative retrieval method
for sponsored search engine, which uses neural machine translation (NMT) to generate keywords
directly from query. This method is completely end-to-end,
which skips query rewriting and relevance judging phases in traditional
retrieval systems. Different from standard machine
translation, the target space in the retrieval setting is a constrained
closed set, where only committed keywords should be generated. We
present a Trie-based pruning technique in beam search to address this
problem. The biggest challenge in deploying this method
into a real industrial environment is the latency impact of running
the decoder.
Self-normalized training coupled with Trie-based dynamic pruning dramatically reduces the
inference time, yielding a speedup of more than 20 times. We also devise an mixed online-offline serving 
architecture to reduce the latency and CPU consumption.
To encourage the NMT to generate new keywords uncovered by the existing system,
training data is carefully selected. 
This model has been successfully applied in Baidu's commercial search
engine as a supplementary retrieval branch, which has brought a remarkable revenue improvement of more
than 10 percents. 
\end{abstract}

%

\ccsdesc[300]{Information systems~Probabilistic retrieval models}

\keywords{sponsored search, keyword retrieval, generative model, beam
  search, Pruning, information retrieval, neural machine translation}

\maketitle

\section{Introduction}  
Sponsored search is an interplay of three entities. The
\emph{advertisers} provide business advertisements and 
bid on keywords to target their ads. The \emph{search engine}
provides the platform where the advertisers' ads can be shown to the
\emph{user} along with organic results. The \emph{user} submits queries
to the search engine and interacts with ads.

Historically, search engine only provides \emph{exact match} type between
queries and keywords. In this scenario, an ad can only be shown when a user's
query exactly matches one of the keywords that the advertiser bids. This puts a great
burden to advertisers, since they have to carefully select hundreds of
thousands of relevant keywords for their business. 

Modern sponsored search platform usually supplies \emph{advanced
  match} type to
release the advertisers from this heavy work of choosing  keywords. In this scenario, keywords
are no longer required to be the same as queries, but should be
semantically relevant to queries. For its simplicity and flexibility, 
\emph{advanced match} type has been
becoming more and more popular among advertisers, and now it accounts
for a large part in search engine's revenue.


Under the condition of advanced match type, query keyword matching is implemented
as a standard information retrieval process, where keyword candidates are
retrieved from an inverted index structure, then a <query, keyword> relevance
model \cite{Hillard_improvingad} would be used to get
rid of low relevant keywords. As is common knowledge, one of the
fundamental problems in information retrieval is the semantic gap between queries
and documents. In sponsored search scenario, the doc's role is embodied
by the keyword. Most ad keywords are short texts, which
increases ambiguity and makes the gap even more serious.

\begin{figure}
  \centering
  \resizebox{.8\totalheight}{!}{
      \includegraphics[width=0.5\textwidth]{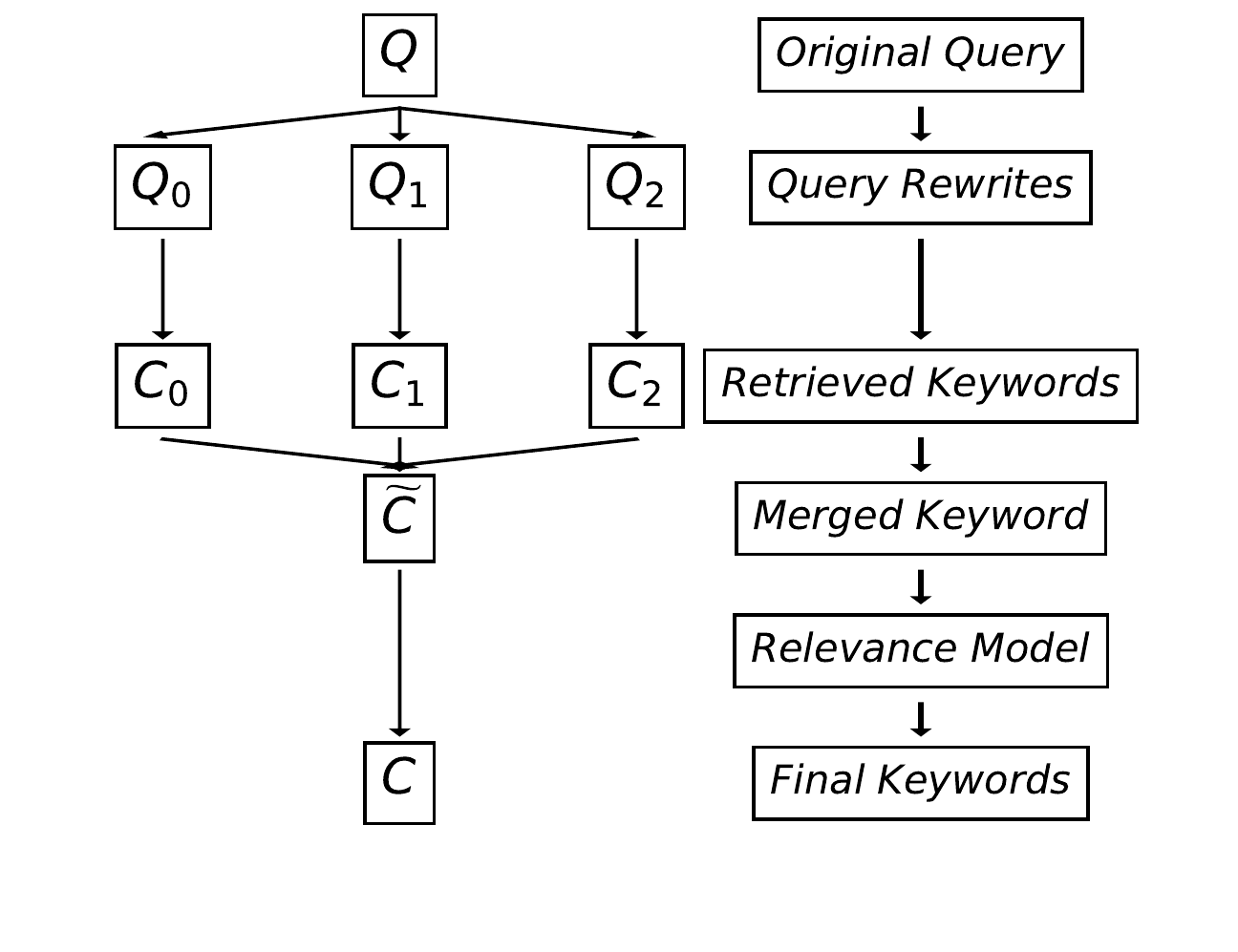}
}
  \caption{Stages of keyword retrieval in sponsored search.}
  \label{fig:keyword_retrieval}
\end{figure}

Most sponsored search systems use query rewriting
technique \cite{Malekian:2008} to alleviate this problem, within which several
query rewrites would be used as alternative queries to retrieve keywords.
As is illustrated in Figure \ref{fig:keyword_retrieval}, the keyword retrieval process usually comprises three stages:
\begin{enumerate}
\item A query $Q$ is rewrited into $L$ sub-queries
  $Q_1, Q_2, \ldots, Q_L$
\item Each sub-query $Q_i$ is submitted to a Boolean retrieval engine
  to get its corresponding candidate list $C_i$, all of the candidates are merged together as
  $\widetilde{C}=\bigcup\limits_{i=1}^L C_i$
\item The candidate set $\widetilde{C}$ would be filtered by a relevance
  judge model to get final keyword set $C$.
\end{enumerate}
A big disadvantage of this framework is the accumulation of
errors. Each sub-module in this framework might
have a trade-off between precision and recall, and also a trade-off
between effect and latency performance. Following the
retrieval path, these errors would be accumulated gradually, and
resulting in a low precision and recall rate finally. 

Monolingual statistical machine translation has been used as a typical method to
generate query rewriting \cite{18Riezler, 8Gao, 14Jones}. With the fast
development of DNN, end-to-end neural machine translation (NMT)
\cite{bahdanau2014} has achieved
a translation performance comparable to the existing state-of-the-art
phrase-based systems \cite{koehn2003statistical}. Recently,
He et al. \cite{learning_to_rewrite_query} applied a sequence-to-sequence LSTM architecture to
rewriting model.

Compared with statistical machine translation (SMT), one great advantage of NMT is that
the whole system can be easily and completely constructed by learning
from data without human involvement. Another major advantage of NMT is that
the gating mechanism (like LSTM \cite{hochreiter1997long}, GRU
\cite{cho2014learning} et al.) and attention
 techniques \cite{bahdanau2014} were proved to be
effective in modeling long-distance dependencies and complicated
alignment relations in the translation process, which posed a serious
challenge for SMT \cite{wang2017neural}. 

\begin{figure}
  \centering
 \resizebox{.4\totalheight}{!}{
     \includegraphics[width=0.5\textwidth]{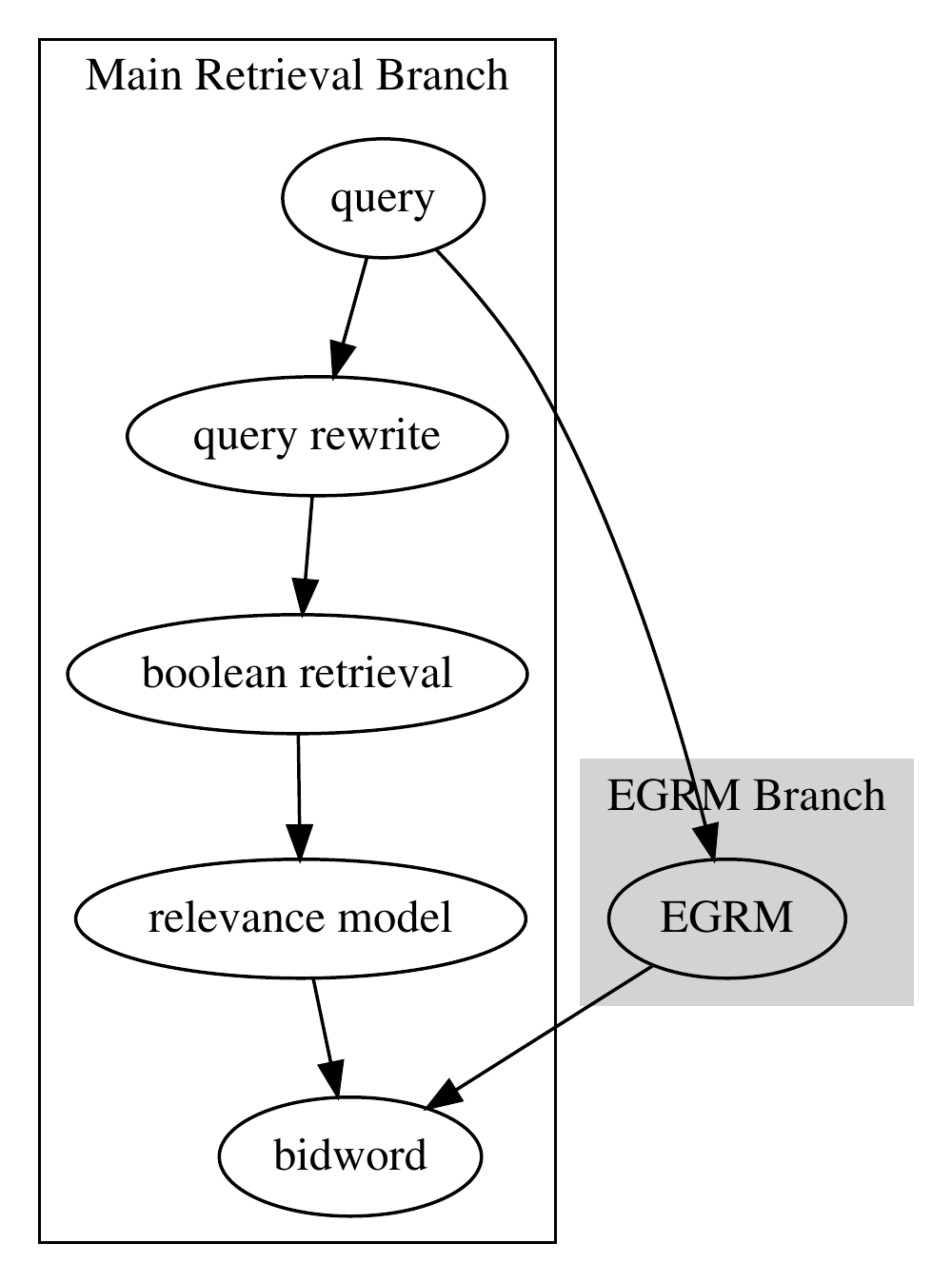}
}
  \caption{The EGRM framework, which skips query rewriting and
    relevance model, is implemented as a separate
    complementary branch to the existing retrieval trunk.}
  \label{fig:trigger_branch}
\end{figure}

Inspired by the aforementioned works, we propose a new retrieval
method named \emph{end-to-end Generative Retrieval Method} to narrow the query keyword
semantic gap, which uses NMT to directly
generate keyword from query. As is illustrated in Figure
\ref{fig:trigger_branch}, the EGRM is implemented as a supplement branch to the existing retrieval
system. To address the error accumulation problem, the query rewriting
and relevance judging phases have been skipped.

\begin{figure}
  \centering
    \includegraphics[width=0.5\textwidth]{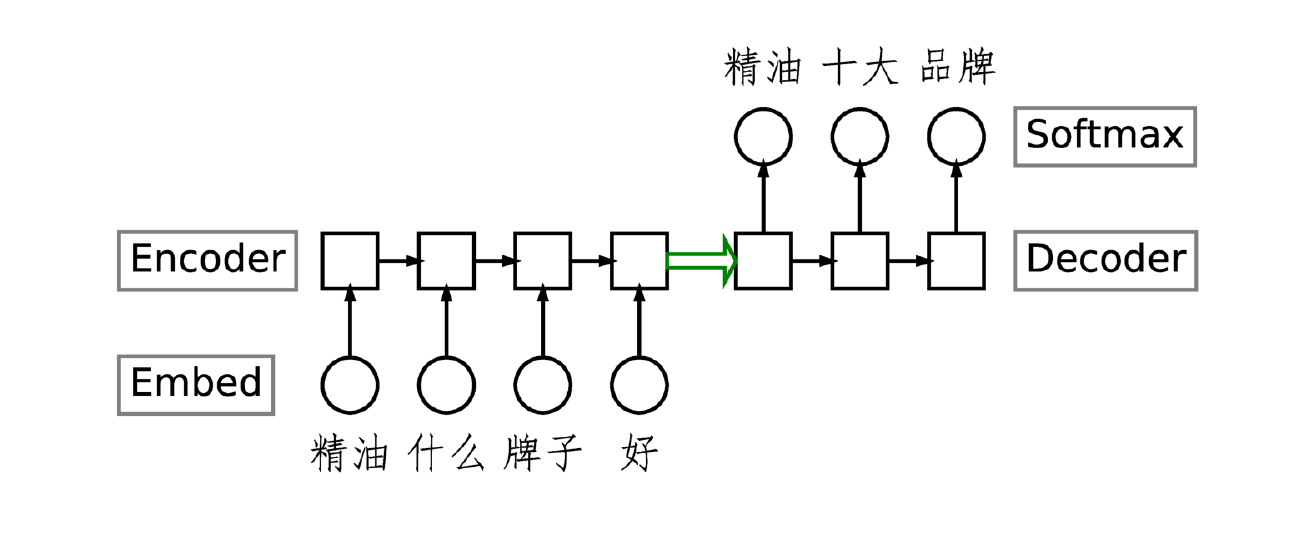}
  \caption{A schematic diagram of the EGRM model structure. A query is
  encoded into a list of hidden states, and then a decoder is used to
generate tokens one by one based on the previous generated tokens and
these hidden states.}
  \label{fig:encoder_decoder}
\end{figure}

Figure \ref{fig:encoder_decoder} is a schematic diagram of the EGRM
model structure. A standard encoder-decoder neural machine translation structure has been
deployed, within which a query is encoded by a multi-layer 
residual Recurrent Neural Network (RNN) encoder into a
list of hidden states, and then a multi-layer residual RNN decoder is
used to decode the target keyword one token by one token based on
these hidden states and the previously
generated tokens. During inference, a beam search strategy is used to
approximately  generate top $k$ best translations.

To carry out this idea in a real industrial environment is a
challenging task.

The biggest challenge is the efficiency of NMT's decoding. 
Standard beam search, which is only able to translate
about ten words per second \cite{hu2015improved}, can hardly meet the requirement
of commercial systems. The average response time for a commercial
sponsored search system is about 200ms.

Secondly, general machine translation is an open target domain
 problem, where there are no restrictions added to the generated
 sentence.  However, decoding in sponsored search scenario
 is an constrained closed target domain problem, where only keywords committed by advertisers are permitted during the generation.

Thirdly, general machine translation focus on generating one best
translation for a source input. In our scenario, we want the
translation model to generate as much unretrieved keywords as possible.  
Here unretrieved keywords refer to the keywords that can not be retrieved
by the current keyword retrieval system. 
 Retrieval in sponsored search is like a link prediction problem in a
 bipartite graph, where queries and keywords are two kinds of nodes,
 and retrieval relationship makes the edges. The more new edges we
 establish, the more ads supply we can make for the downstream auction queue.
 As a supplement to the current retrieval system, the EGRM framework is
 encouraged to trigger more unretrieved keywords.

Our key contributions in this work are the following:
\begin{enumerate}
\item An end-to-end generative retrieval method is introduced in
  sponsored search engine, which skips query rewriting and relevance
  judge model. This framework has been successfully implemented in
  Baidu's commercial search engine, which has contributed a revenue 
improvement of more than 10\%. To our knowledge, this is the
first published job of applying NMT as a generative retrieval model
into sponsored search engine. We hope this would shed light on
  further design of sponsored retrieval system and NMT's
  application in industry.
\item A Trie-based pruning technique is introduced into the beam
  search, which greatly solved the constrained target domain problem.
\item Self normalization accompanied with Trie-based dynamic pruning dramatically
  reduced the decoding time, which yields a speedup of more than 20
  times.
\item We carefully selects the organic log results to encourage the NMT to
  generate more unretrieved keywords.
\end{enumerate}

\section{Related Work}

Machine translation is a popular way to alleviate the
semantic gap in the NLP domain. With parallel corpus, machine translation can learn the underlying
word alignment between target words and source words. If we use
monolingual parallel data, semantic synonymy can be
detected. Basically, there are two kinds of applications of machine
translation in information retrieval. 
The first one uses machine translation as a discriminative model to evaluate
<query, doc> relevance. 
Given a query $Q$ and a document $D$, the translation
probability $P(D|Q)$ or $P(D|Q)$  was used as a feature
to boost the calculation of query document relevance
\cite{yahoo_webrank, micro_gao_clickthrough}.
 \citet{Hillard_improvingad}  applied this idea to calculating the commercial query ad relevance.
The second one uses machine translation as a generative model to
directly generate relevant candidates. This idea has been widely used
in query rewrite. 
\citet{18Riezler, 8Gao, 14Jones} treated query rewrite as a
statistical machine translation problem with monolingual training
data. 
Recently, \citet{learning_to_rewrite_query} proposed a sequence to sequence
deep learning framework to study the query rewrite.

The most related work to ours is the paper recently published by \citet{Lee_2018}, 
which used conditional GAN to generate keywords from queries. 
There are several critical points that make our work different from theirs:
\begin{itemize}
\item The target domain in their translation setting is not closed. The
  generated sentence might not be a valid keyword.
\item Unlike their approaches, we do not include commercial click log in our training data. This allows the NMT to generate  more words not covered by the existing system.
\item Our work concentrates on addressing the latency impact of
  deploying the generative model into the real commercial
  system. Nevertheless, they  showed no experiment results in the
  industry environment.
\end{itemize}

Although NMT gives us a nice and simple end-to-end way to deploy a state-of-the-art
machine translation system, its decoding efficiency is still
challenging. The standard beam search algorithm implemented by  \citet{bahdanau2014}
reduced the search space from exponential size to polynomial size, and
is able to translate about ten words per second . However,
this speed is still far from 
meeting our requirement of commercial online retrieval
systems. \citet{hu2015improved} built a priority queue to further reduce
the search space. And they also introduced a constrained softmax operation
which uses phrase based translation system to generate the constrained
word candidates. Since lots of unnecessary hypothesis are removed,  the
computational efficiency is greatly improved. 

\section{Background}

In following formulas, we use bold lower case to denote
vectors(e.g. $\bold{q}$), 
capital letters to represent
sequences(e.g. $Q, K$), squiggle letters to represent
set(e.g. $\mathcal{K}$) and lower case to represent individual tokens
in a sequence (e.g. $q_1, k_2$), $k_{<i}$ to represent the token
sequence $k_0, k_1, \ldots, k_{i-1}$, where $k_0$ is a special beginning of sentence symbol that is
prepended to every target keyword.

Let $(Q, K)$ be a <query, keyword> pair, where
$Q=q_1,q_2,\ldots,q_M$ is the sequence of $M$ tokens of source query $Q$, 
and $K=k_1,k_2,\ldots, k_N$ is the sequence of $N$ tokens in the target
keyword $K$. 
From the probabilistic perspective, machine translation is equivalent to maximizing
the log likelihood of the conditional probability of sequence $K$ given a source query $Q$, i.e.,
$\log P(K|Q)$, which can be
decomposed into factors:
\begin{equation}
  \label{eq:chain_rule}
  \log P(K|Q) = \sum\limits_{i=1}^N \log P(k_i|k_{<i};Q)
\end{equation}

Our model follows the common
sequence to sequence learning encoder-decoder framework
\cite{Sutskever2014} with attention \cite{bahdanau2014}. 
Under this framework, an encoder reads the input query
$Q$ and encode its meaning into a list of hidden vectors:
\begin{equation}
  \label{eq:encoder}
  \bold{Q}=(\bold{q}_1,\bold{q}_2,\ldots, \bold{q}_M)=\mathrm{Encoder}(q_1, q_2, \ldots, q_M)
\end{equation}
where $\bold{q}_i\in \mathbb{R}^n$ is a hidden state at time $t$. 
In our experiment, the encoder is mainly implemented by RNN:
\begin{equation}
  \label{eq:RNN}
  \bold{q}_i = \mathrm{RNN}(q_i, \bold{q}_{i-1})
\end{equation}

And the decoder is trained to predict the probability 
of next token $k_i$ given the
hidden states $\bold{Q} = (\bold{q}_1,\bold{q}_2,\ldots, \bold{q}_M)$ and all the
previously predicted words $k_1,\ldots, k_{i-1}$
\begin{equation}
  \label{eq:chain_rule0}
P(k_i|k_{<i};Q) \approx P(k_i|k_{<i};\bold{Q}).
\end{equation}
During inference, target tokens would be decoded one by one 
based on this distribution, until a special end of sentence symbol(<e>) is generated.

In order to focus on different parts of the
source query during decoding, an attention mechanism \cite{bahdanau2014} is introduced to
connect the decoder hidden states and encoder hidden states. Let
$k_{i-1}$ be the decoder output from the last decoding time step $i-1$, 
$\bold{c}_i$ be the attention context for the current time step $i$,
which is calculated according to the following formulas:
\begin{equation}
  \label{eq:attention}
  \begin{split}
\bold{c}_i & = \sum\limits_{j=1}^{M} \alpha_{ij} \bold{q}_j, \\
\alpha_{ij} &= \frac{\mathrm{exp}(e_{ij})}{\sum\limits_{p=1}^M\mathrm{exp}(e_{ip})}, \\
e_{ij} &= \mathrm{Atten}(\bold{k}_{i-1}, \bold{q}_j)
  \end{split}
\end{equation}
where $\mathrm{Atten}$ could be implemented as dot product or feed forward
network and $\bold{k_i}$ is the hidden state vector at time step $i$.

The RNN decoding phase is computed as follows:
\begin{equation}
  \label{eq:softmax}
  \begin{split}
\bold{k}_i &= \mathrm{RNN}(\bold{k}_{i-1}, k_{i-1}, \bold{c}_i), \\
p(k_i=w|k_{<i};Q) &= \frac{\mathrm{exp}(s_i(w))}{\sum_{w'}\mathrm{exp}(s_i(w'))} \\
s_{i}(w) &= s(k_{i-1}, \bold{k}_i, \bold{c}_i, w)
  \end{split}
\end{equation}
where $s_i(w)$ is the unnormalized energy score of choosing $k_i$ to
be $w$.


\section{Selecting training data: Difference Oriented}
\label{sec:select_dataset}

As a complementary branch to the main retrieval system,
linking underlying unretrieved relevant query keyword together is our
major concern. We hope the most keywords generated by EGRM are
unretrieved ones, especially considering that the decoding
phase takes a lot of time.

Click logs are used as parallel corpus for training the NMT.
Typically, there are two kinds of click logs in commercial search engine, the
organic click log and sponsored ads click log. Sponsored ads log provides
commercial (query-keyword) click pairs which are also the current
retrieval system's feedback. Using the feedback looped log as training data would not 
generate much difference, since maximum likelihood estimation
would make the top $k$ keywords to be the same as those in the training data.
 Organic click log provides us with natural (query-title)
click pairs. The vast difference between organic and paid search
results makes it possible for the NMT to generate more keywords different
from the current retrieved ones. 

\section{Decoding efficiently into a closed set}
\label{sec:closed_set}
One challenge in applying machine translation to keyword retrieval
task is that our target space is a restricted fixed set  of submitted keywords,
whereas in general translation, the target space is unconstrained, 
which means any possible sentences might be generated.

There are several possible methods to mitigate this problem. 
Firstly, we might generate as many candidates as possible, then pick out the real keywords. 
However, this is not applicable in a low latency industrial environment, since decoding much
candidates would cost a lot of time.

Secondly, we might use <query, keyword> data from commercial
click log as our training data. Translation is essentially a
conditional language model. A language model trained with 
<query, keyword> data is supposed to guide the decoder to generate real keywords.
However, as is pointed out in the last section, this would
not induce much difference to our current retrieval system.

In this paper, we devise a novel pruning technique in beam search
called Trie-based pruning to fix this problem.


\begin{figure}
  \centering
    \includegraphics[width=0.5\textwidth]{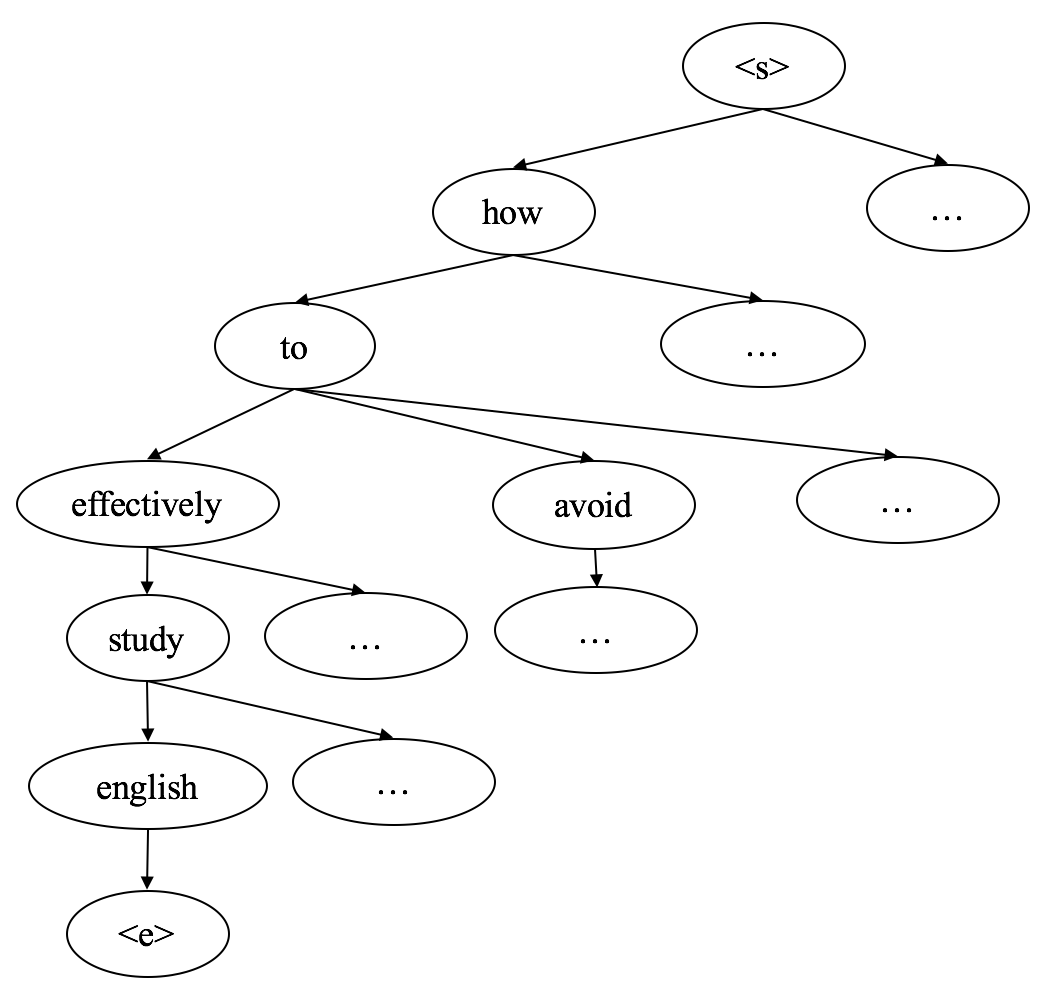}
  \caption{A prefix tree is built before the decoding phase.}
  \label{fig:prefix_tree}
\end{figure}
A prefix tree $T_{\mathcal{K}}$ for the constrained keyword set
$\mathcal{K}$ is built before the decoding phase. 
Fist of all, each keyword $K$ in $\mathcal{K}$ is tokenized. Then we use
these token lists to build a prefix tree keyed by tokens as is illustrated in
Figure \ref{fig:prefix_tree}. 


\subsection{Trie-based pruning within a beam search}
\begin{figure}
  \centering
    \includegraphics[width=0.5\textwidth]{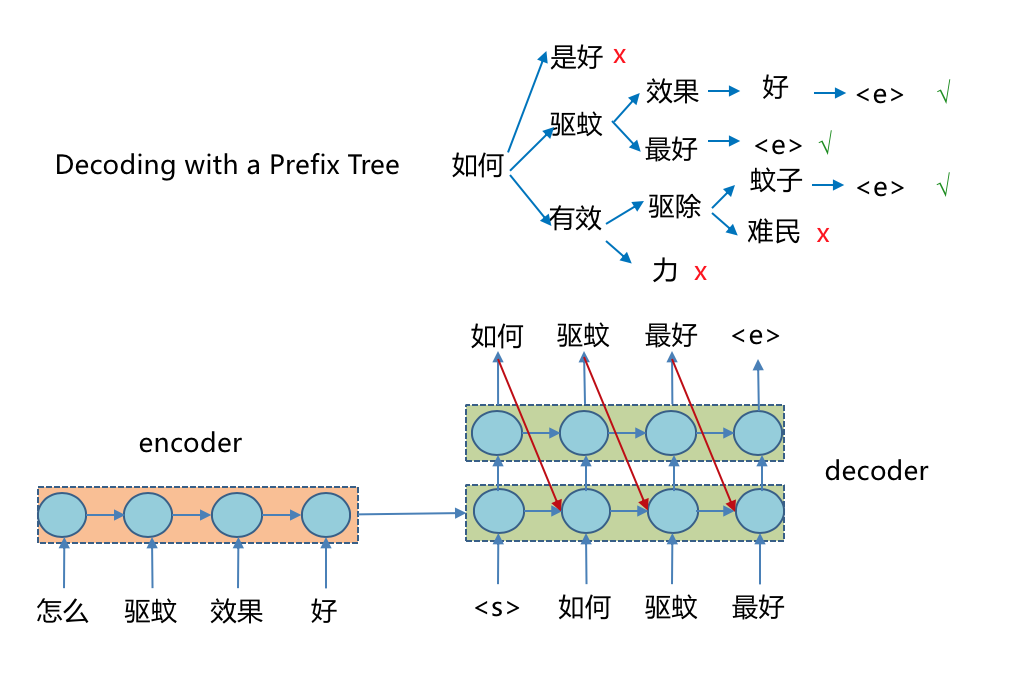}
  \caption{Trie-based pruning within a beam search. At the inference stage, we beam search with a prefix tree
   to decode the query into a closed bidword set on-the-fly. Specifically, at each step of the beam search,
   the prefix tree will directly give the valid suffix tokens following the current hypothesis path,
   After that, a greedy top-k selection is performed within the legal tokens. In this way we make sure
   all generated sentences are committed bidwords and computation time is saved at the meantime. }
  \label{fig:beamsearch}
\end{figure}

Suppose we are doing a beam search of size $B$, at the $i$th 
decoding phase, $i-1$ tokens have been generated, $B$ hypotheses are
conserved in the following set $\mathrm{BeamSet}=\{\hat{K}^j_{i-1}=(k^j_1,
k^j_2,\ldots, k^j_{i-1}), 1\leq j\leq B\}$. For each hypothesis
$\hat{K}^j_{i-1}$, 
the model would inference a conditional token probability of
$p(k^j_i|k^j_1, \ldots, k^j_{i-1})$. With a prefix
tree $T_{\mathcal{K}}$, we can get all the valid suffix tokens set $\mathrm{suffix}(\hat{K}^j_{i-1})$ directly following
the trie path $\hat{K}^j_{i-1}$, then only the valid
suffix in $\mathrm{suffix}(\hat{K}^j_{i-1})$ would be kept, other tokens
would be pruned away. Figure \ref{fig:beamsearch} shows the whole pruing process.
With the Trie-based pruning technique, all the generated sentences 
are valid keywords, which greatly improves efficiency.

Another important feature of using Trie-based pruning is that: only a
small limited number of tokens in the large vocabulary need to be calculated. In fact, 
Table \ref{tab:prefix_tree_layer_length} shows the average suffix
token numbers for each layer of the prefix tree, which is built for
295 billion keywords. Going from the top
to bottom, the suffix token number decreases quickly, which makes it possible
to gain a great speedup with Trie-based pruning technique.
%
%
\begin{table}[h]
\centering
\begin{tabular}{cc}
\toprule
\textbf{Layer} & \textbf{Suffix Number} \\
\midrule
0 & 38090 \\
1& 297.75 \\
2 &4.97\\
3 & 2.28\\
4 & 1.54 \\
5 & 1.23\\
6 & 1.17\\
7 & 1.20\\\bottomrule

\end{tabular}
\caption{The average suffix numbers for top 7 layers of the prefix tree.}
\label{tab:prefix_tree_layer_length}
\end{table}

\subsection{Self-normalization}
It is well known that one serious performance bottleneck at inference stage is 
the computation of the denominator of the softmax, i.e. $\sum_{w'}\mathrm{exp}(s_i(w'))$ 
in equation \ref{eq:softmax}, as it involves summarization over the entire output vocabulary space. 
Various approaches have been proposed to address this problem \cite{ruder2016softmax}. 
Inspired by the balanced binary tree, \citet{morin2005hierarchical} proposed to replace the 
flat softmax layer with a hierarchical tree. Recently
\citet{grave2017efficient} came up with adaptive softmax for efficient
computation on GPU, which handles frequent words and 
infrequent words separately with different hidden state sizes. 
Another kind of approach is the sampling-based. \citet{bengio2003quick} proposed Important Sampling
 to reduce the computation. \citet{mikolov2013distributed} used
 Negative Sampling to address the problem. More sophisticated methods like Noise  
Contrastive Estimation \cite{mnih2012fast} are also available.

Following \citet{Devlin2014FastAR}'s work, we use the self normalizing trick to speed up the
decoding. To be specific, during training, an explicit regularization loss is added to the original
likelihood loss in Equation \ref{eq:chain_rule} to encourage the softmax normalizer to be as close
to 1 as possible. 
\begin{equation}
  \label{eq:self-normalize}
  \begin{split}
    L = &\sum\limits_{i} \log(P(k_i|k_{<i}, Q)) - \beta(\log(\sum_{w'}\mathrm{exp}(s_i(w')))-0)^2 \\
    = &\sum\limits_{i} \log(P(k_i|k_{<i}, Q)) - \beta(\log(\sum_{w'}\mathrm{exp}(s_i(w'))))^2 \\
  \end{split}
\end{equation}

When decoding with self-normalized model, the costly step for calculating the
denominator  $\sum_{w'}\mathrm{exp}(s_i(w'))$ is avoided, we only have to compute the numerator
$s_i(w)$. 

Furthermore, combined with a prefix tree, a small number of numerators need to
be calculated. As a matter of fact, we can just predict the valid suffix
words conditioned on the current output words path, which would save
much more time. 

\subsection{Drop inferior hypotheses on the fly}
Another useful trick in our implementation is to remove inferior hypotheses
on the fly. 
Generally, a likelihood threshold is set up to filter the final
generated keywords at the end of decoding. This threshold can also be
used in the internal process of decoding. 
As we decode a new token based on the current hypothesis,
the likelihood of hypotheses would be multiplied by anther probability
factor, therefore the full likelihood decreases as decoding proceeds.
Based on this consideration, if the current hypothesis's likelihood is lower than
the given threshold, we would not expand it out later. This trick would make more qualified keywords(with a likelihood
above the threshold) in the final generated hypothesis set. Combined
with the Trie-based pruning, the total decoding time would also be decreased.

\subsection{Full Algorithm}
The full algorithm is shown as
in Algorithm \ref{alg:algo}.
\begin{algorithm} [h]
    \caption{Beam Search with Trie-based pruning} 
    \KwIn{Beam Size $B$, Self-normalized NMT $M$, Keyword Prefix tree
        $T$, score threshold $s_{min}$} 
    \KwOut{Keywords set $Out$} 
        cur\_buffer $\leftarrow \emptyset$ \; 
        next\_buffer $\leftarrow \emptyset$ \;
        $Out \leftarrow \emptyset$ \;
        put <s> into cur\_buffer\;
        \While{{\rm{cur\_buffer is not empty and size$(Out)< B$ }}}
        { 
            \For {{\rm{each hypothesis $c$ in cur\_buffer}}}
            {
              get the valid suffix word set $S_c$ for $c$ with $T$\;
              \For{{\rm{each suffix word $w$ in  $S_c$}}}
              {
                extend partial hypothesis $c$ with $w$ to get new hypothesis $\tilde{c}=[c;w]$\;
                using $M$ to inference score $s_{\tilde{c}}$ for $\tilde{c}$\;
                \If{$s_{\tilde{c}} > s_{min}$} 
                {
                  \If{$w$ == {\rm{<e>}} }
                  {
                    put $\tilde{c}$ into $Out$
                  }
                  \Else
                  {
                    put $\tilde{c}$ into next\_buffer\;
                  }
                }
              }
            } 
            sort elements $\tilde{c}$ in the next\_buffer according to their
            score $s_{\tilde{c}}$ and keep only the top $B -$ size$(Out)$. \;
            cur\_buffer $\leftarrow$ next\_buffer\;
            next\_buffer $\leftarrow \emptyset$ \;
          }
        return $Out$ \; 
     \label{alg:algo}
    \end{algorithm}


\subsection{An online-offline mixing architecture}
\begin{figure}
  \centering
    \includegraphics[width=0.5\textwidth]{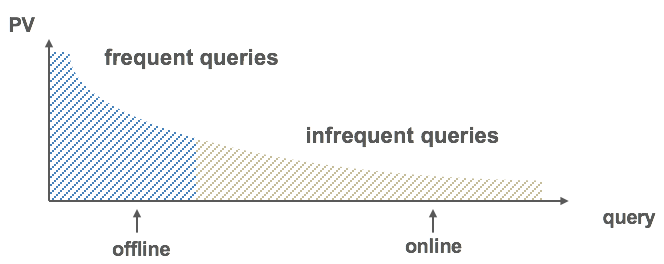}
  \caption{An online-offline mixing architecture: Keywords for
    frequent queries are generated completely offline with a complex
    model, whereas keywords for infrequent queries are generated
    online with a simple model.}
  \label{fig:online_offline}
\end{figure}

Large commercial search engines report ever-growing query volumes, leading to
tremendous computation load on sponsored ads.
It is well known that these search queries are highly skewed and exhibit a power law distribution
\cite{powerlaw2001, Petersen2016}. That is, in a fixed time period, the most popular queries compose the head
and torso of the curve, in other words, approximately 20\% of queries
occupies 80\% of the query volume. This property has motivated the
cache design upon search results. Inspired by this idea, we designed a
online-offline mixing architecture in Figure \ref{fig:online_offline}.

Under this framework, query volume is divided into two parts: frequent
queries and infrequent queries. For frequent queries, their generated keywords
are computed completely offline, where enormous computing resources
can be used. In our experiment, we deployed a complex model with a 4
layer LSTM encoder and a 4 layer LSTM decoder; for infrequent queries,
keywords are generated online, where latency is strictly restricted.
In this scenario, we implemented a simple model which is a single
layer GRU Gated RNN encoder and a single layer GRU Gated
RNN decoder. This mixed framework helps us to save more than 70\% cpu resources.

\section{Experiments}
In this section, we conduct experiments to show the performance of our
proposed EGRM framework.

\subsection{Training Data Set}
As mentioned in section \ref{sec:select_dataset}, in order to encourage the NMT to
generate more unretrieved keywords,  we include organic user click log instead of commercial user click log in out training data, where the latter one is the feedback of current commercial retrieval system.
749 million query-title pairs are sampled from
Baidu's one month user click log. Titles are simply prepossessed to
trim the last domain name related part. 
Queries and titles are tokenized, and top frequent tokens are kept
to form the vocabulary. Other words are all mapped into the same UNK
token. Our vocabulary size is 42,000.

Table \ref{tab:basic_statistic} shows some basic statistics of the
data. There are 3.5 tokens in query, 6.5 tokens in title and 4.5
tokens in keyword on average. The prefix tree is built for 295 million \emph{advanced match} type keywords.

\begin{table}[h]
\centering
\begin{tabular}{lrc}
\toprule
\textbf{Field} & \textbf{Size}(million) &  \textbf{Average Length} \\
\midrule
Query & 749  &  5.01 \\
\midrule
Title & 749  &  6.22 \\
\midrule
Keyword & 295 &  4.68 \\
\bottomrule
\end{tabular}
\caption{The statistics of the training data.}
\label{tab:basic_statistic}
\end{table}

\subsection{Implementation Details}
We use Adam \cite{kingma2014adam} with Xavier weight initialization
\cite{glorot2010understanding} 
as the optimizer  to perform Stochastic Gradient Ascent(SGA).
The initial learning rate is set to be $5\times10^{-4}$ and
the mini-batch size is 128. The hidden state vectors' dimension is 512.

The offline model is implemented with a
four layer LSTM encoder and a four layer LSTM decoder with attention.
And the online model is implemented as a one layer GRU Gated RNN and a one
layer GRU gated decoder with attention. Self-normalization and Trie
based dynamic pruning have been applied in both online and offline situations.
We use paddlepaddle \footnote{http://www.paddlepaddle.org/} as the DNN
training and inference tool. The Trie-based pruning strategy
is fully realized with C++ language.

\subsection{Decoding Efficiency}
We describe our experimental settings as follows. 
\emph{Baseline} is the typical sequence to sequence GRU Gated RNN with
a one layer encoder and a one layer decoder, and the decoding is
realized with standard beam search. 
'SN' means training with self-normalization and 'TP' means decoding with Trie-based pruning. 'DropOTF' refers to the strategy of dropping inferior hypotheses on-the-fly.

All experiments are conducted on our EGRM server. 10,000 queries 
are randomly sampled from Baidu's commercial engine log and used as
input to the EGRM system.

Table \ref{tab:speedup} shows the decoding time preformed with
different strategies and different beam sizes. As is seen from this
table, 'SN + TP' strategy
greatly reduces the decoding time, reaching a speedup of nearly 10
times. Combined with 'DropOTF', the decoding time can be further decreased by nearly
one half. 
\begin{table}[h]
\centering
\begin{tabular}{cccc}
\toprule
\textbf{Beam Size}  &  \tabincell{c}{\textbf{Baseline}} & \tabincell{c}{\textbf{Baseline}\\ \textbf{+SN+TP}} & \tabincell{c}{\textbf{Baseline+SN+TP}\\ \textbf{+DropOTF}} \\
\hline
    40	&443.40 &45.48	 &25.64 \\\midrule
    60	&663.22 &66.03	 &28.43 \\\midrule
    80	&853.05 &86.42	 &30.66 \\\midrule
    100	&1125.09& 123.49 &		32.15 \\
\bottomrule
\end{tabular}
\caption{The average decoding time(in millisecond) of different strategies in different beam size.}
\label{tab:speedup}
\end{table}

\subsection{Validity of Generated Hypotheses without Trie Pruning}
\begin{figure}[h]
  \centering
    \includegraphics[width=0.5\textwidth]{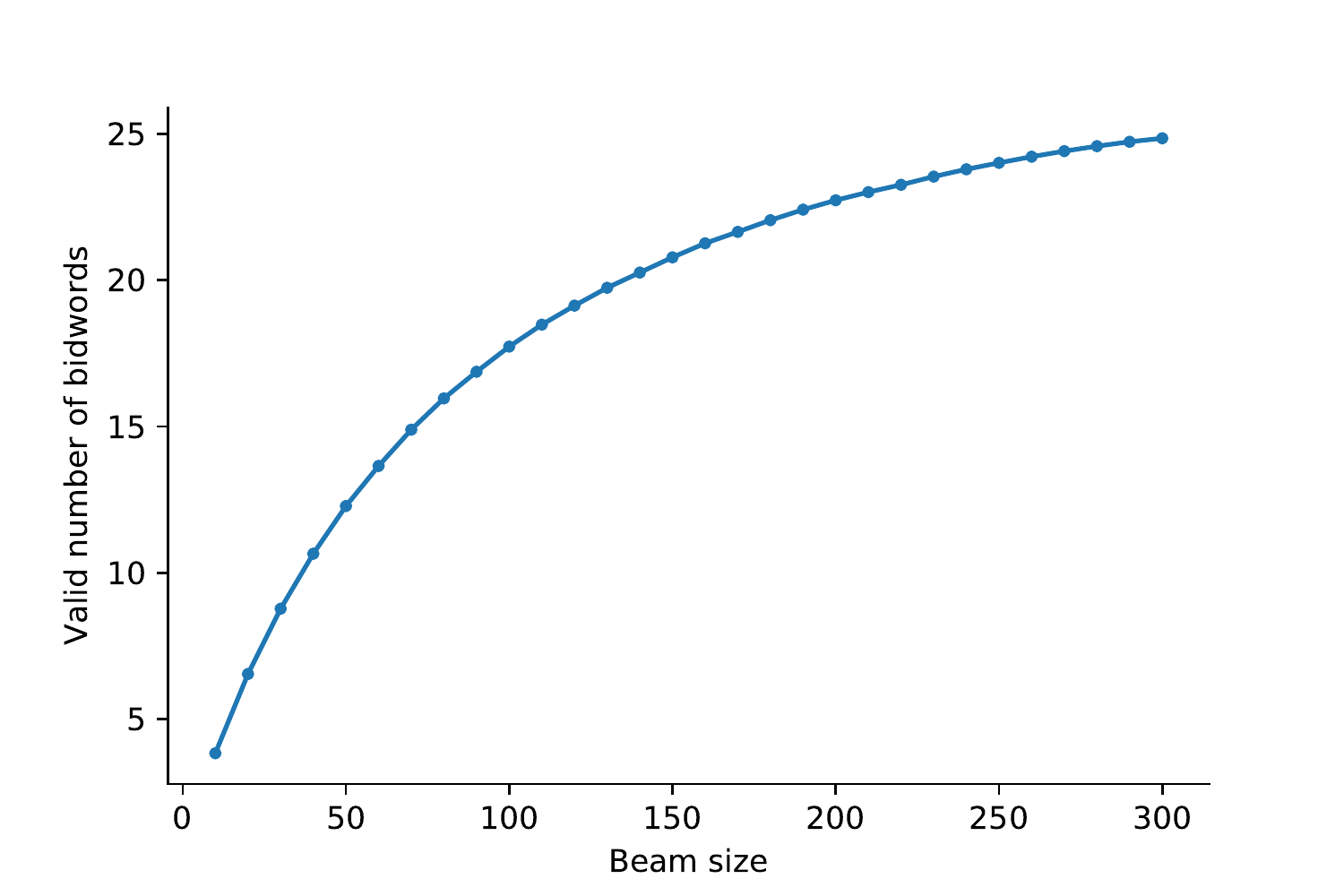}
  \caption{When Trie-based  pruning is removed, only a small number of
    the beam search results are actual keywords.}
  \label{fig:valid_keyword_proportion}
\end{figure}

The following experiment shows the  necessity of using \emph{Trie
 based Pruning} to decoding into a closed set. As mentioned in section
\ref{sec:closed_set}, using \emph{Trie-based Pruning} can garantee
that all generated sentences are valid keywords. 
Figure \ref{fig:valid_keyword_proportion} shows that when Trie-based pruning is
removed, only a small number of the generated sentences are actual keywords.  As
the beam size increases, the amount of valid keywords increased quite
slowly. To be specific, when the beam size is set to be as large as 300, only 8\%
of the results are valid keywords.

\subsection{Relevance Assessment}
Query-keyword pairs are sampled from online A/B test experiment, with
800 pairs sampled on each individual side. These pairs are sent to professional human
judges for three grade labels: good, fair and bad. Table \ref{tab:judge} shows the A/B judgment for generated keywords
and baseline results. For commercial privacy concern, we only show the relative
improvements based on the current system. As is shown in Table
\ref{tab:judge}, the bad case proportion has dropped by
-20.7\% compared with the existing system's, and the good case
proportion has increased by 6.6\%. This demonstrated that: under the
condition of a great speedup, our EGRM system can still generate high quality keywords.
\begin{table}[]
  \centering
\centering
\begin{tabular}{cc}
\toprule
\textbf{Quality}  &  \textbf{Improvements}\\
\midrule
Good & +6.6\%\\
\midrule
Fair & +3.1\%\\
\midrule
Bad & -20.7\%\\

\bottomrule
\end{tabular}
  \caption{Relevance comparison is conducted between the EGRM generated keywords and
  the current system's retrieved keywords.}
  \label{tab:judge}
\end{table}

\subsection{Online Evaluation}
We also conduct online experiment for our EGRM system with real traffic.
We use two metrics to evaluate the performance of our retrieval framework.
\begin{enumerate}
\item CTR denotes the average click ratio received by the search
  engine, which can be formalized as
  $\frac{\#\{\rm{clicks}\}}{\#\{\rm{searches}\}}$(one search means one
  submit of a query).
\item CPM denotes revenue received by search engine for 1000
  searches, which can be formalized as $\frac{\rm{revenue}}{\#\{\rm
    {searches}\}}\times 1000$.
\end{enumerate}

As is shown in Table \ref{tab:online_evalu},  the EGRM system has
contributed to a CPM growth of 13.8\%  and a 
CTR growth of 15.4\%. 
These metrics demonstrate that
our EGRM branch has a better semantic understanding, and it does
create a significant mount of new links for the underlying relevant
query-keyword pairs. 
\begin{table}[h]
  \centering
\centering
\begin{tabular}{cc}
\toprule
\textbf{Quality}  & \textbf{Improvements}\\
\midrule
CPM & +13.8\% \\
\midrule
CTR & +15.4\% \\
\bottomrule
\end{tabular}
  \caption{Online A/B Test performance of the EGRM system.}
  \label{tab:online_evalu}
\end{table}


\section{Conclusions}
In this paper, we have proposed a novel generative
retrieval method for sponsored search engine. This method is fully
 end-to-end, without using query rewriting and relevance model. 
To make the decoded sentences limited within a closed domain,  a
Trie-based pruning mechanism has been introduced. 
To meet the real-time interaction demand of sponsored ad system, we
have introduced self-normalization training technique coupled with
dynamic Trie-based pruning. Experiments have demonstrated that our model can reduce the
generating time to one twentieth without degrading the relevance quality.
In addition, training data has been carefully selected to encourage
the NMT to generate unretrieved keywords. Further, taking advantage of the power
law distribution of queries, a mixed online-offline architecture has
been constructed to save the CPU resources. 

\section{Future Works}
We believe that decoding into a constrained domain is not a specific
problem only suited for keyword retrieval. 
For example, task-oriented dialogue systems \cite{gao2018neural} might
be required to generate or retrieve
answers within a closed set, e.g. music name or lyrics. For the purpose of safe search,
we might also want to limit the generation of certain phases and the prefix tree trick
can help filter them on-the-fly. 

To further improve the decoding efficiency, we could build several
prefixed trees. When a query is submitted, a trade classifier would
predict its trade, then a keyword prefix tree in the same trade is
chosen. Finally, decoding would be restricted on this prefix tree.
Since trades provide a natural boundary to link query and keywords. Queries and keywords should not be
linked across trades. 

\bibliographystyle{ACM-Reference-Format}
\bibliography{Bibliography-File}

\end{document}